\documentclass[journal]{IEEEtran}
\usepackage{graphicx} 
\usepackage{float} 
\usepackage{subfigure} 
\graphicspath{ {./images/} }
\usepackage{stfloats}
\usepackage{color,xcolor}
\usepackage{makecell}
\usepackage{booktabs}
\usepackage{multirow}
\usepackage{algorithm}
\usepackage{algpseudocode}
\usepackage{amsmath,amssymb}  
\usepackage{bm,esint}
\usepackage{tabulary}
\usepackage{cuted}
\usepackage{makecell}
\usepackage{amsfonts} 

\algnewcommand{\StateNoWrap}[1]{%
  \State \makebox[0pt][l]{#1}%
}

\definecolor{bg1}{RGB}{177,130,0}\colorlet{bg1}{bg1!50}
\definecolor{bg2}{RGB}{0,85,179}\colorlet{bg2}{bg2!50}
\definecolor{bg3}{RGB}{0,134,6}\colorlet{bg3}{bg3!50}
\definecolor{bg4}{RGB}{161,154,7}\colorlet{bg4}{bg4!50}
\definecolor{bg5}{RGB}{245,83,158}\colorlet{bg5}{bg5!50}
\ifCLASSINFOpdf
\else
\fi

\usepackage[]{hyperref}
\usepackage[hyperref=true,backend=biber,sorting=none,backref=false]{biblatex}
\bibliography{citizen}

\usepackage{tikz,xcolor,hyperref}
\definecolor{darkgreen}{RGB}{0,100,0} 
\definecolor{lime}{HTML}{A6CE39}
\DeclareRobustCommand{\orcidicon}{%
 \begin{tikzpicture}
 \draw[lime, fill=lime] (0,0) 
 circle [radius=0.16] 
 node[white] {{\fontfamily{qag}\selectfont \tiny ID}}; \draw[white, fill=white] (-0.0625,0.095) 
 circle [radius=0.007]; \end{tikzpicture}
 \hspace{-2mm}}
\foreach \x in {A, ..., Z}{%
 \expandafter\xdef\csname orcid\x\endcsname{\noexpand\href{https://orcid.org/\csname orcidauthor\x\endcsname}{\noexpand\orcidicon}}
 }

\begin{document}
%
\title{PoisonCatcher: Revealing and Identifying LDP Poisoning Attacks in IIoT}
%
%
%

\author{Lisha~Shuai\orcidA{},
 Shaofeng~Tan\orcidB{},
 Nan Zhang,
 Jiamin~Zhang\orcidC{},
 Min Zhang\orcidD{},
 Xiaolong~Yang\orcidE{}
 
\thanks{This work was supported by the National Natural Science Foundation of China under Grant 61971033. (\textit{Corresponding author: Xiaolong~Yang}.)}
\thanks{Lisha Shuai, Shaofeng Tan, Jiamin Zhang, Min Zhang and Xiaolong Yang are with School of Computer and Communication Engineering, University of Science and Technology Beijing, Beijing, 100083, China (e-mail: shuails@xs.ustb.edu.cn; winber@126.com; zhangjm\_ustb@163.com; zhangm@ustb.edu.cn; yangxl@ustb.edu.cn).}
\thanks{Nan Zhang is with the State Grid Information \& Telecommunication Group Co., Ltd., Beijing, 101200, China (e-mail: 837885930@qq.com).}
\thanks{Shaofeng Tan is with the Information Management Center, Beijing Pinggu Hospital, Beijing, 101200, China.}
}

%
%

\markboth{Journal of \LaTeX\ Class Files,~Vol.~14, No.~8, August~2015}%
{Shell \MakeLowercase{\textit{et al.}}: Bare Demo of IEEEtran.cls for IEEE Journals}
%



\maketitle

\begin{abstract}

Local Differential Privacy (LDP), a robust privacy-protection model, is widely adopted in the Industrial Internet of Things (IIoT) due to its lightweight, decentralized, and scalable. However, its perturbation-based privacy-protection mechanism hinders distinguishing between any two data, thereby facilitating LDP poisoning attacks. The exposed physical-layer vulnerabilities and resource-constrained prevalent at the IIoT edge not only facilitate such attacks but also render existing LDP poisoning defenses, all of which are deployed at the edge and rely on ample resources, impractical.

This work proposes a LDP poisoning defense for IIoT in the resource-rich aggregator. We first reveal key poisoning attack modes occurring within the LDP-utilized IIoT data-collection process, detailing how IIoT vulnerabilities enable attacks, and then formulate a general attack model and derive the poisoned data's indistinguishability. This work subsequently analyzes the poisoning impacts on aggregated data based on industrial process correlation, revealing the distortion of statistical query results' temporal similarity and the resulting disruption of inter-attribute correlation, and uncovering the intriguing paradox that adversaries' attempts to stabilize their poisoning actions for stealth are difficult to maintain. Given these findings, we propose PoisonCatcher, a solution for identifying poisoned data, which includes time-series detectors based on temporal similarity, attribute correlation, and pattern stability metrics to detect poisoned attributes, and a latent-bias feature miner for identifying poisons. Experiments on the real-world dataset indicate that PoisonCatcher successfully identifies poisoned data, demonstrating robust identification capabilities with F2 scores above 90.7\% under various attack settings.

\end{abstract}


%
\IEEEpeerreviewmaketitle

\section{Introduction}

The Industrial Internet of Things (IIoT) revolutionizes industrial operations and drives advances across sectors. By integrating physical infrastructure with the digital realm, IIoT enables real-time monitoring, efficient management, and comprehensive analysis of data, facilitating decision making, improving productivity, and boosting cost-efficiency \cite{khan2020industrial}. However, the ongoing evolution of the IIoT drives frequent data sharing among numerous devices and systems, which, due to the inherent complexity and potential vulnerabilities within these interconnected environments, leads to increased exposure to sensitive and valuable information, thereby increasing the risk of attacks on the IIoT infrastructure and assets. Such risks can result in production disruption, reputation damage, and threats to community safety and national security. Therefore, protecting data privacy and ensuring secure data-sharing are fundamental requirements for the sustained growth and stability of IIoT.

Local Differential Privacy (LDP) \cite{kasiviswanathan2011can}, a robust privacy protection model, is widely employed in IIoT due to its lightweight, independence, and scalability \cite{wang2019local}. By adding independent and identically distributed (i.i.d.) noise, LDP provides strong privacy guarantees for individual data while maintaining acceptable statistical query result (SQR) accuracy \cite{duchi2013local}. Tech giants leverage LDP in data collection, enhancing data privacy while obtaining aggregated insights crucial for operations and improvements. For instance, Microsoft (Windows 10) \cite{ding2017collecting}, Xiaomi (MIUI), and Meizu (Flyme) use LDP to collect geolocation data. Google employs LDP to gather Chrome usage statistics \cite{erlingsson2014rappor}, while Apple uses it to collect emojis and word usage \cite{2017LearningWP}. LDP is also crucial in sectors such as smart grids \cite{shanmugarasa2023local}, healthcare \cite{arachchige2019local} and manufacturing \cite{jiang2021differential}.

However, data perturbation in LDP, enacted for privacy, inherently complicates distinguishing clean from poisoned data \cite{cheu2021manipulation}. Exploiting this indistinguishability, adversaries can effectively launch data poisoning attacks against LDP-utilized systems (hereafter referred to as LDP poisoning attacks) \cite{272214}. While LDP offers some resilience against minor poisoning, IIoT’s unique complexity exacerbates the problem. Specifically, high-volume real-time data streams strain processing capabilities, and the resource limitations imposed by the 3C (Communication, Computation, and Caching) constraints of industrial sensors hinder the deployment of advanced security measures. These factors collectively create a fertile ground for LDP poisoning attacks. Consequently, the intricate interplay between LDP indistinguishability and IIoT complexity necessitates tailored countermeasures.

Therefore, this study examines the LDP vulnerabilities in IIoT. We first reveal three key attack modes, formulating a generalized attack model that illustrates their stealth properties. Analysis shows how such attacks distort statistical query results (SQR), consequently disrupting intrinsic inter-attribute correlations, and reveals a critical characteristic, that is, the attack pattern rarely remains stable over time. Drawing upon this finding, we propose PoisonCatcher, a countermeasure that identifies poisoned data. PoisonCatcher integrates three complementary detectors for monitoring SQR changes and detecting poisoned attributes, and a latent-bias feature miner for identifying specific poisoned data. Our key contributions are as follows:
\begin{itemize}
\item Revealing vulnerability in LDP-utilized IIoT, where adversaries can exploit the interaction between LDP's indistinguishability and IIoT's vulnerabilities to launch poisoning attacks, undermining statistical accuracy and consequently compromising decision-making reliability.
\item Recognizing three key LDP poisoning attack modes specific to IIoT, and crucially, deriving and analyzing their three main impacts, which provide a solid foundation for designing robust countermeasures.
\item Developing PoisonCatcher, a novel defense engineered for poisoned data identification, addresses the challenges faced by the resource-constrained IIoT edge in defending against LDP poisoning attacks.
\item Validating PoisonCatcher's effectiveness in a real-world dataset, achieving an F2 score exceeding 90.7\% for poison identification across various conditions.
\end{itemize}

\section{Related Work}

Recent studies have revealed the severe impact of LDP poisoning attacks on data utility. Cao et al. \cite{272214} and Wu et al. \cite{279934} proved that LDP poisoning attacks significantly compromise the accuracy of the frequency and mean estimation. Particularly, Wu et al. further verified that in key-value data scenarios, adversaries could substantially affect SQRs through carefully crafted false data. Cheu et al. \cite{cheu2021manipulation} revealed that even minor poison could inject significant bias into various statistic estimations, with effects exacerbated under low privacy budgets or large data collection domains. Li et al. \cite{li2022fine} investigated the impact mechanisms on the estimation of mean and variance, while Imola et al. \cite{cheu2022differentially} and Cheu et al. \cite{imola2022robustness} explored how these attacks increase the estimation errors of histograms and their relationship with the intensity of the attack and privacy budgets. These studies not only elucidated the severity of LDP poisoning attacks but also guided subsequent defense mechanism design.

The development of countermeasures against LDP poisoning attacks has evolved from isolated breakthroughs to comprehensive defense strategies. In 2021, two complementary technical directions emerged. Cao et al. \cite{272214} proposed fundamental defense mechanisms that include normalization, conditional probability detection, and fake user identification at the level of the LDP protocol. The verifiable LDP mechanism based on the cryptographic random response technique (CRRT) proposed by \cite{10.1007/978-3-030-81242-3_3} combines theoretical guarantees with an encryption approach to enhance security verification. This method embeds verifiable elements in the data randomization process, allowing the aggregator to validate whether clients have followed the prescribed randomization protocols, thereby advancing cryptographic applications in the defense of LDP.

During 2022-2023, researchers have increasingly focused on improving the efficiency and universality of the defense mechanism. Wu et al. \cite{279934} proposed detecting fake user behavior patterns like data anomalies and communication repetition to counteract poisoning attacks. Although partially effective, it struggles with parameter tuning, false positives, and high computational costs. The interaction-based verifiable LDP protocol \cite{10220122} inherited and improved the verification concepts of CRRT, reducing computational overhead through collaborative perturbation mechanisms, while the emPrivKV protocol \cite{10.1007/978-3-031-33498-6_17} combined expectation maximization algorithms with secure random sampling, improving defense universality while maintaining the advantages of the earlier encryption approach.

Research works in the year 2024 demonstrated concurrent theoretical breakthroughs and practical innovations. The LDPGuard framework \cite{10415225} achieved unified defense against three types of attack and LDP protocols through innovative two-round collection methods. Horigome et al. \cite{app14146368} complemented emPrivKV by integrating encryption protocols with expectation maximization algorithms, further improving defense mechanisms in key-value data scenarios. Zheng et al. \cite{10423870} modeled attacks as double-layer optimization problems from a theoretical perspective, providing mathematical foundations for defense mechanism design. Masahiro et al. \cite{10.1007/978-3-031-68208-7_18} investigated the application in an oblivious transfer (OT) protocol to enhance the robustness of the LDP Count Mean Sketch (CMS) against poisoning attacks. To address the increased data transmission and processing costs with longer CMS vectors, they introduced Hadamard CMS using the Hadamard transform.

However, existing works exhibit notable limitations. Basic defense mechanisms, such as those in \cite{272214}, show insufficient detection accuracy against complex attack patterns and struggle to adapt to dynamic attack strategies. Cryptography-based methods (\cite{10.1007/978-3-030-81242-3_3}), while providing verifiability guarantees, increase computational and communication costs, hindering efficient deployment in large-scale distributed systems. Recent comprehensive frameworks (\cite{10.1007/978-3-031-33498-6_17, 10415225}), despite advances in universality, remain dependent on prior knowledge of attack patterns and show significant performance degradation when processing high-dimensional data. Moreover, existing methods focus on client verification or protocol improvements, overlooking the potential of aggregator data distribution characteristics for attack detection.

Consequently, leveraging statistical perspectives from the aggregator is of significant importance to defend against LDP poisoning attacks. This approach inherently offers lower computational complexity, which makes it particularly advantageous for the 3C resource-constrained IIoT edge by avoiding an additional burden on edge nodes and facilitating real-time defense scenarios. More crucially, by analyzing the statistical characteristics of the aggregated results, it can overcome the reliance on prior attack patterns and effectively handle high-dimensional data and dynamic attack strategies. Thus, statistics-based approaches can complement and strengthen existing defense strategies, contributing to a more robust LDP defense system.

\section{LDP Mechanisms and Vulnerabilities in IIoT}

While widely adopted for IIoT data privacy, LDP can inadvertently obscure malicious activities. This duality requires a comprehensive understanding of its privacy benefits and security drawbacks in IIoT. Hence, this section explores LDP's theoretical foundations, advantages, and vulnerabilities from indistinguishability and IIoT complexities.

\textit{LDP Theoretical Foundations}: Let $X_i \in \mathcal{X}$ be the input and $Z_i \in \mathcal{Z}$ be the output corresponding to the LDP on $\mathcal{X}$. For any $x, x' \in \mathcal{X}$ and any measurable subset $S \subseteq \mathcal{Z}$, given a privacy budget $\varepsilon > 0$, if a statistical query function $\mathcal{Q}$ satisfies the following inequality, then $Z_i$ is said to be an $\varepsilon$-LDP representation of $X_i$ \cite{duchi2013local}:
\begin{equation}
    \sup_S \frac{\mathcal{Q}(Z_{i} \in S | X_{i} = x)}{\mathcal{Q}(Z_{i} \in S | X_{i} = x')} \leq e^\varepsilon
    \label{LDP_definition_1}
\end{equation}
where $\mathcal{Q}(Z_i \in S \mid X_i = x)$ denotes the conditional probability that the output $Z_i$ falls within the set $S$, given the input $X_i = x$, denoted as $\mathbb{P}(Z_i \in S \mid X_i = x)$. Each output $Z_i$ depends solely on its corresponding input $X_i$, represented as $X_i \rightarrow Z_i$, and $Z_i$ is independent of all other inputs and outputs given $X_i$, expressed as $Z_i \perp \{X_j, Z_j \mid j \neq i\} \mid X_i$. This independence is crucial for LDP implementation, as it limits the threat of leaking information about inputs through the output $Z_i$.

\textit{LDP Advantages}: The unique benefits of LDP make it an ideal choice to protect sensitive data in the IIoT. By performing data perturbation locally, it eliminates the need for trusted central collectors or third parties, thereby mitigating the threats of large-scale data breaches. Its simple randomization techniques not only ensure data privacy, but also reduce computational load on 3C resource-constrained edge devices. The flexible and adjustable privacy levels allow for optimal balancing of protection and utility based on specific security needs.

\textit{LDP Vulnerabilities in IIoT}: LDP's inherent indistinguishability, while crucial for data privacy, presents a major vulnerability to poisoning attacks. Adversaries can exploit this to inject malicious data into LDP-processed datasets, affecting the accuracy in SQR and the reliability of critical decisions. This threat is particularly pronounced in IIoT, where dynamic topologies complicate anomaly detection, massive data volumes can conceal poisoned samples, and the 3C resource-constrained edge limits robust security measures, making it a prime target for data poisoning. Moreover, IIoT-specific factors further amplify this vulnerability. Increased physical access to devices increases the risk of tampering, while heterogeneous ecosystems with inconsistent security offer multiple attack vectors. The criticality of industrial processes also magnifies the impact of even subtle data poisoning. Ultimately, these IIoT complexities significantly exacerbate the vulnerabilities and potential consequences of LDP poisoning attacks. 

\section{LDP Poisoning Attacks in IIoT}

To bridge the gap in understanding the LDP poisoning attack in IIoT, this section analyzes their underlying mechanisms and impacts. We begin by revealing key poisoning attack modes and examining how IIoT vulnerabilities contribute to their execution. Following this, we formulate a general attack model and analyze the stealthiness characteristics of poisoned data. The section then analyzes the impacts of such attacks on aggregated data, leveraging industrial process correlation. This analysis highlights key findings, including the distortion of SQRs and the disruption of attribute correlation, as well as an intriguing paradox regarding attack stability. Together, the insights lay the groundwork for designing countermeasures. The key symbols and their definitions are outlined in Table \ref{table: symbol and description}.

\begin{table}[htpb]
\centering
\caption{Symbol and Description}
\begin{tabular}{c>{\centering\arraybackslash}m{3cm}p{3.7cm}}
 \hline
 Symbol  &Mathematical Expression& Description\\ 
 \hline
 $n$  &$ n \in \mathbb{N}  ^ { + }$& Number of industrial device using LDP\\
 $T$& $ T \in \mathbb{N}  ^ { + }$&Total number of discrete time instances\\
 $t$  &$ t \in \left\{ 1 , 2 , \ldots , T \right\}$& A discrete time instance\\ 
 $k$  &$ k \in \mathbb{N}  ^ { + }$& Number of attributes in the dataset\\
 $d_{ij}$&$ d_{ij} \in \mathbb{R}$ or other domain& Clean data of the $j$-th attribute for the $i$-th devices \\
 $D_j$&$D_j=\left [ d_{1j},\cdots,d_{nj} \right ] \in \mathbb{R}^n$& Clean single-attribute dataset for the $j$-th attribute\\ 
 $\mathcal{D}^{\text{hist}}$ &Set of datasets& Set of historical clean datasets\\
 $\mathcal{X}$ &Domain, e.g., $\mathbb{R}  ^ { k }$& Data collection and output domain\\
 $\Delta$  &Function / Mechanism& Poisoning function applied by adversaries\\
 $\Phi$  &Function / Mechanism& Standard function in the absence of poisoning attacks\\
 $\psi $ &Algorithm / Mechanism& Perturbation algorithm satisfying $\varepsilon$-LDP\\ 
 $\varepsilon$  &$\varepsilon > 0$& Privacy budget for LDP\\ 
 $ \delta $  &$ 0 \leq \delta < 1 $& Confidence probability\\
 $\gamma$  &$\gamma \geq 0$& Additional error bound \\
 $\mathcal{Q}$ &Function& Statistical query function  (e.g., mean, frequency)\\ 
 $ \Lambda $ &$ \Lambda \geq 0$& limited variation or maximum discrepancy\\
\hline
\end{tabular}
\label{table: symbol and description}
\end{table}

\subsection{LDP Poisoning Attack Modes in IIoT}

Previous studies have classified LDP poisoning attacks from different perspectives. Cheu et al. \cite{cheu2021manipulation} proposed a binary framework based on attack phases, i.e., input poisoning attack (IPA), which compromises data collection, and output poisoning attack (OPA), which involves tampering during transmission. Cao et al. \cite{272214} presented a three-category framework centered on attack strategies, i.e., (i) Random Perturbation Attack (RPA), similar to input poisoning; (ii) Random Item Attack (RIA), similar to output poisoning; and (iii) Maximum Gain Attack (MGA), a targeted and strategic attack. In IIoT settings, however, beyond exhibiting unique characteristics of existing poisoning modes, we identify and formalize a novel attack mode that extends the theoretical framework.

The inherent security limitations of IIoT create a diverse type of attack surface. First, industrial devices are deployed in open physical environments with exposed hardware interfaces, enabling adversaries to inject malicious code via physical layer tampering. Second, strict end-to-end low-latency requirements in industrial communication constrain the deployment of computationally intensive security mechanisms, thereby creating a time window at the transport layer that adversaries exploit for attacks. Third, industrial networking protocols typically only verify transmission integrity while failing to validate data semantics, creating exploitable vulnerabilities for attacks. Finally, legitimate variations in LDP implementation parameters and the absence of real-time validation across IIoT devices provide effective cover for malicious modifications, enabling coordinated cross-device attacks.

\begin{figure}[htbp]
\centering 
\includegraphics[width=\linewidth]{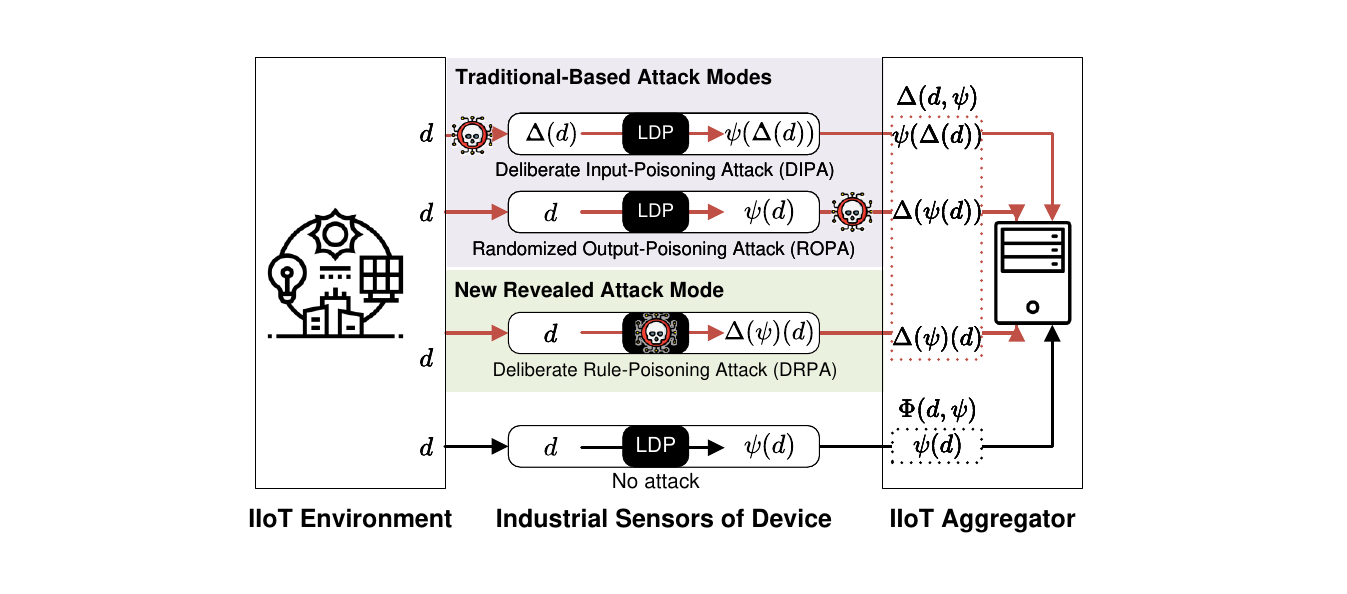}
\caption{Three Distinct Attack Modes of the LDP Poisoning Attack in IIoT}
\label{LDP Poisoning Attack}
\end{figure}

Therefore, integrating the attack classifications from Cheu et al. and Cao et al. with insights into the unique characteristics and vulnerabilities of IIoT, we identify three distinct LDP poisoning attack modes based on the key stages of the LDP-utilized IIoT data-sharing process, as shown in Fig. \ref{LDP Poisoning Attack}.

\subsubsection{Deliberate Input-Poisoning Attack (DIPA)}

An IIoT-specific instance combining RIA and IPA, which distorts the statistical distribution by poisoning the sensing samples. Such poisoning is facilitated through techniques that include, but are not limited to, electromagnetic interference, deliberate alteration of environmental parameters, or hijacking data at the firmware level. Adversaries compromise the utility of sensor readings by leveraging the deficiency in self-verification mechanisms within industrial sensors, alongside inherent vulnerabilities present at the physical layer of IIoT access protocols.

Due to the indistinguishability of the LDP-processed data, it is difficult to detect deliberate and random poisoning. However, adversaries seeking to manipulate the SQR will utilize deliberate poisoning, as random attacks are uncontrollable and inefficient to achieve their goals. This attack mode transforms an authentic reading $d$ into a falsified $ \Delta(d) $. Applying a $\varepsilon$-LDP mechanism $ \psi_\varepsilon $ then yields $ \psi_\varepsilon(\Delta(d))$, which is statistically inseparable from the legitimate output $ \psi_\varepsilon(d) $. Formally, the attack mode is:
\begin{equation}
\Delta_{\text{in}}(d, \psi_\varepsilon) = \psi_\varepsilon(\Delta(d)), \quad\text{s.t. } \Delta(d) \in R_{\text{sen}} \cap R_{\text{env}} \label{stealth_1}
\end{equation}
This constraint is imposed to ensure the stealth of the attack, where $R_{\text{sen}}$ denotes the sensor's physical measurement range and $R_{\text{env}}$ represents the valid range based on the deployment environment.

\subsubsection{Deliberate Rule-Poisoning Attack (DRPA)}

A new attack mode specific to IIoT settings proposed by us. In the IIoT, the lack of rigorous validation of privacy parameters in the 3C resource-constrained edge devices, combined with firmware vulnerabilities that allow direct injection or modification of rule sets, allows adversaries to initiate DRPA. By poisoning LDP rules, typically through compromised firmware updates or insecure remote configuration protocols, adversaries can compromise both the integrity and the intended functionality of LDP mechanisms deployed on edge devices (e.g., sensors, actuators, and PLCs).

Specifically,  DRPA poisons the LDP mechanism by: (i) adjusting the privacy parameters within the reasonable range to manipulate the noise distribution; (ii) modifying the perturbation mechanism to introduce asymmetric noise; and (iii) reconstructing the data encoding and mapping functions to distort the data representations. Such poison may undermine the guarantees of privacy-utility trade-off (e.g., unbiasedness) inherent in the original LDP design.

Formally, the devices apply a poisoned LDP mechanism $\Delta(\psi_\varepsilon)$ to their authentic reading $d$, ensuring that the result conforms to the $\varepsilon$-LDP guarantee and is statistically indistinguishable from legitimate results. The generated output, $\Delta(\psi_\varepsilon)(d)$, is then sent to the aggregator. The attack mode is defined as:
\begin{equation}
    \Delta_{\text{rule}}(d, \psi_\varepsilon) = \Delta(\psi_\varepsilon)(d) , \quad\text{s.t. } \begin{matrix}\sum_{i =1}^{n}\end{matrix} \varepsilon_{i} = \varepsilon_{\text{total}}\label{stealth_2}
\end{equation}
where $n$ represents the number of devices employing the LDP. DRPA can introduce targeted systematic bias into aggregated statistics as long as the poisoned implementation passes detection systems that solely verify formal compliance indicators, such as preserving the invariant total privacy budget.

\subsubsection{Randomized Output-Poisoning Attack (ROPA)}

An instance of RPA combined with OPA is for IIoT scenarios. Its feasibility is enhanced by two common IIoT limitations, i.e., stringent low-latency requirements that preclude intensive output validation, and communication protocols that primarily verify transmission integrity rather than data semantics. Exploiting these limitations, adversaries can initiate ROPA to inject statistically subtle, randomized noise into the data output space. The reason is that a simple deterministic modification of the LDP output is readily detectable, often due to factors like limited output space or the LDP mechanism's specific output distribution pattern. Consequently, adversaries are compelled to employ randomization to modify the output, thereby emulating LDP noise characteristics and achieving stealth.

The adversary samples a new value, denoted $ \Delta(\psi(d)) $, from a subset $ \mathcal{X} \subseteq \text{Range}(\psi) $ that comprises the outputs deemed legitimate by the encoding mechanism. While individual data under $\varepsilon$-LDP inherently prevent adversaries from identifying whether a specific record has been perturbed, continuous data collection in IIoT introduces unique security challenges. Specifically, an adversary who compromises a device and perpetually poisons its long-term data streams must ensure that each malicious perturbation adheres to the prescribed LDP mechanism to evade detection. Formally, the attack mode is defined as \eqref{stealth_3}.

\begin{figure*}[ht]
  \centering
  \begin{equation}
    \Delta_{\text{out}}(d, \psi_\varepsilon) = \Delta(\psi_\varepsilon(d)) \quad\text{s.t. } \mathbb{P}_{\mathcal{X}}\bigl( \Delta(\psi_\varepsilon(d^t)) = x \bigm| d^1, \dots, d^{t-1} \bigr) \propto \exp\left( -\frac{\varepsilon \cdot \| x - \psi_\varepsilon(d^t) \|_1}{\Delta f}\right)\label{stealth_3}
  \end{equation}
\end{figure*}

\noindent{where $\Delta f$ denotes the sensitivity of the query function. This rule serves as the constraint that ensures stealth for long-term attacks on a single device. It enables the stream of poisoned outputs to statistically mimic legitimate LDP perturbed data over time, thereby preserving the appearance of privacy compliance, achieving undetectability in continuous IIoT.}

\subsection{Theoretical Framework for LDP Poisoning Attacks}

This section formulates a comprehensive general formulation that unifies three LDP poisoning attack modes and derives the mathematical indistinguishability of poisoned data, providing a foundation for quantifying their statistical impact.

\subsubsection{Generalized Formulation of LDP Poisoning Attacks}

Despite operating at different stages of the LDP-utilized IIoT data-sharing process, these poisoning attack modes all induce a systematic distortion of the established $\varepsilon$-LDP guarantee. The standard $\varepsilon$-LDP guarantee, particularly when applied to a statistical query function $\mathcal{Q}$, aims to limit the risk of inferring individual information from mechanism outputs, as formally defined in \eqref{LDP_definition_1}. However, poisoning attacks replace the original mechanism $\Phi$ with a corrupted mechanism $\Delta$. The change in privacy characteristics induced by such attacks can be quantified as follows:
\begin{equation}
\sup_{\substack{d, d' \in \mathcal{X},  S \subseteq \mathcal{Z}}} \frac{\mathbb{P}[\Phi(d,\psi) \in S]}{\mathbb{P}[\Delta(d',\psi) \in S]} \leq e^{\varepsilon + \varepsilon'}
\label{revised_general_bound_final}
\end{equation}
where $\varepsilon'$ is the distortion magnitude caused by the attack.

Each attack mode operates under specific constraints while manipulating its contribution to this privacy deviation $\varepsilon'$ to maintain stealth. For DIPA, a single attack instance achieves $\varepsilon'=0$ under constraint \eqref{stealth_1}, rendering its output indistinguishable from legitimate noise. This stealth enables the long-term cumulative effect, leading to a gradual deviation of the device-level data distribution and affecting the integrity of the SQR. DRPA maintains stealth by enforcing privacy budget balance, specified by constraint \eqref{stealth_2}. This balance is achieved either globally across compromised devices ($\sum_{i \in M} \varepsilon'_i = 0$) or temporally for long-term attacks on a single device ($\sum_{t=1}^T \varepsilon'^{t}_i = 0$). ROPA, operating through malicious post-processing, adheres to an exponential decay distribution property as specified by constraint \eqref{stealth_3}. This involves manipulating the output probabilities based on the $L_1$ distance to the true value, thereby maintaining stealth.

Crucially, the primary objective of such attacks is not to maximize the leakage of individual privacy information, but rather to systematically poison the dataset by manipulating the data generation process, thereby disrupting the effectiveness of SQR. Through a carefully designed mechanism $\Delta$, the adversary causes the probability distribution of the output $\tilde{Z}_i = \Delta(X_i, \psi)$ to shift. Consequently, when the data aggregator applies a statistical query $\mathcal{Q}$ to these attacked outputs $\{\tilde{Z}_i\}$, even if the query logic itself remains unchanged, the final aggregated statistical results (e.g., mean, frequency distribution, etc.) will significantly deviate from the true or expected results based on the original data $\{X_i\}$ or unattacked outputs $\{Z_i\}$. This deviation directly affects the statistical utility of the data.

\subsubsection{Indistinguishability Properties of Poisoned Data\label{Poisoned Data Points are Stealthy}}  

This indistinguishability provides significant stealth for poisoned data against anomaly detectors, as noted by Cheu et al. \cite{cheu2021manipulation}. Using the generalized formalization in \eqref{revised_general_bound_final}, we formally demonstrate this property below.

Based on \eqref{revised_general_bound_final}, for any anomaly detector $ \mathcal{I}: \mathcal{X} \to \{0,1\} $, the detection probability difference is:  
\begin{equation} \label{eq:prob_diff_bound_revised}  
\begin{aligned}  
&\left| \mathbb{P}(\mathcal{I}(\Phi) = 1) - \mathbb{P}(\mathcal{I}(\Delta) = 1) \right| \\& \leq \int_{S} \left| \mathcal{Q}(S|\Phi) - \mathcal{Q}(S|\Delta) \right| dS  \leq e^{\varepsilon + \varepsilon'} - 1
\end{aligned}  
\end{equation}

When $ \varepsilon' \to 0 $ (minimal privacy perturbations), the bound collapses to the original LDP guarantee:
\begin{equation}  
\left| \mathbb{P}(\mathcal{I}(\Phi) = 1) - \mathbb{P}(\mathcal{I}(\Delta) = 1) \right| \leq \varepsilon + \varepsilon'
\end{equation}

The stealth probability of poisoned data follows:
\begin{equation}  
\mathbb{P}(\mathcal{I}(\Delta) = 0) \geq \mathbb{P}(\mathcal{I}(\Phi) = 0) - (\varepsilon + \varepsilon')
\end{equation}

At its lower bound where $\varepsilon' = 0$, the attacks result in perfect stealth. Conversely, the upper bound $\varepsilon'_{\text{max}}$ is constrained by the defender's anomaly detection threshold $\varepsilon_{\text{th}}$, requiring that $\varepsilon + \varepsilon'_{\text{max}} \leq \varepsilon_{\text{th}}$ to successfully evade detection mechanisms. The above theoretical bounds formally characterize how adversaries parasitize LDP's privacy-preserving mechanism to launch stealthy poisoning attacks.

\subsection{Impacts of LDP Poisoning Attacks in IIoT}

Industrial processes are characterized by various intrinsic correlations, including temporal correlations (i.e., the dependencies between consecutive measurements over time), spatial correlations (i.e., the dependencies between measurements from physically proximate devices), and attribute correlations (i.e., the dependencies between different physical properties, e.g., current and voltage related by circuit resistance, or temperature readings from nearby sensors related by physical distance). These intrinsic correlations are a prominent feature of industrial data and are susceptible to LDP poisoning attacks.

Building on that, this section first quantitatively analyzes how the LDP poisoning attack distorts the SQR accuracy over the temporal aggregation phases, based on the findings of \cite{cheu2021manipulation}. Crucially, we demonstrate that this distortion consequently disrupts inherent attribute correlations. Observing these two impacts over time-series further reveals a significant extrinsic impact rooted from poisoning, i.e., the attack pattern often exhibits instability. Understanding the interleaved effects between intrinsic correlations and extrinsic impacts is key to characterizing the attack's impact and provides a foundation for developing countermeasures.

\subsubsection{LDP Poisoning Attacks Distort SQR Accuracy  \label{Undermines the Accuracy of Statistical insights}}

Let $N$ be the set of $n$ devices, producing measurements $d_{ij}$ for the attribute $j$ from the device $i$. Assume a subset of devices $M \subseteq N$ is poisoned. The devices $i \in M$ undergo the poisoning operation $\Delta$, while those from the remaining devices $i \in N \setminus M$ are processed by standard perturbation $\Phi$. The dataset $D_j$ collected by the aggregator for attribute $j$, comprising these processed values from all devices, is defined as:
\begin{equation}
\Delta(D_j, \psi) = \bigcup_{M} \{\Delta(d_{ij}, \psi)\} \cup \bigcup_{N \setminus M} \{\Phi(d_{ij}, \psi)\} \label{attack_data}
\end{equation}
which characterizes how the aggregator synthesizes differentially perturbed data streams. Crucially, the coexistence of these asymmetric perturbation mechanisms introduces latent distortion in the aggregation process.

Define the attack‐induced distortion in SQR as the divergence between the estimator’s outputs before and after poisoning. Invoking the Lipschitz continuity of the statistical estimator $\mathcal{Q}$ \cite{heinonen2005lectures}, we obtain an explicit upper bound on this distortion. This continuity, typical of robust estimators, ensures that any bounded perturbation of the input yields a proportionally bounded change in the output of $\mathcal{Q}$, making it particularly amenable to analysis in poisoning attacks that impose restricted data modifications. Let $L>0$ be the Lipschitz constant of $\mathcal{Q}$.  Then, for any poisoned dataset $\Delta(D_j,\psi)$ defined in \eqref{attack_data}, the attack‐induced distortion in SQR is:

\vspace{-10pt}

{\small
\begin{equation}
\|\mathcal{Q}(\Delta(D_j, \psi)) \!-\! \mathcal{Q}(\Phi(D_j, \psi))\|_1 \!\leq\! L \|\Delta(D_j, \psi) \!-\! \Phi(D_j, \psi)\|_1 \!\label{Lipschitz_continuity}
\end{equation}
}

Under the $\varepsilon$-LDP constraint (\ref{LDP_definition_1}), the maximum deviation between any unpoisoned data  $\Phi(d_i, \psi)$ and its poisoned counterpart $\Delta(d_i, \psi)$ is bounded by:  
\begin{equation}
\left\| \Delta(d_i, \psi) - \Phi(d_i, \psi) \right\|_1 \leq e^{\varepsilon} - 1\label{maximum_possible_deviation}
\end{equation}
where this bound stems from the $\varepsilon$-LDP requirement that the probability ratio of any two outputs lies within $[e^{-\varepsilon}, e^{\varepsilon}]$, translating to an additive difference cap of $e^{\varepsilon} - 1$ in the probability domain.  

By integrating this per-device constraint into the Lipschitz continuity framework \eqref{Lipschitz_continuity}, we derive a refined global error bound for the attack scenario defined in \eqref{attack_data} based on \eqref{maximum_possible_deviation}:
\begin{equation}
\begin{split}
&L \left\| \Delta(D_j, \psi) - \Phi(D_j, \psi) \right\|_1 \\
\leq & L \sum_{i \in M} \left\| \Delta(d_{ij}, \psi) - \Phi(d_{ij}, \psi) \right\|_1 \leq L m \left( e^{\varepsilon} - 1 \right)
\end{split}
\label{eq:accuracy_degradation}
\end{equation}

Normalizing by the total dataset size $n$, we find that the distortion is proportional to:
\begin{equation}
    \text{Distortion} \propto L \cdot \frac{m}{n} \cdot \left( e^{\varepsilon} - 1 \right)
\end{equation}
which exposes a critical vulnerability, that is, even with a small poisoned subset, the exponential dependence on $\varepsilon$ enables adversaries to induce significant statistical distortion in SQRs.

\subsubsection{LDP Poisoning Attacks Disrupt Attribute Correlation \label{Compromise the Correlation between Datasets}}  

Targeted perturbations applied to specific attributes give rise to correlation decoherence, whereby both linear and higher‑order dependencies among features are systematically distorted. Consider a dataset with $ k $ attributes, each comprising $ n $ data. Let $ D_x = (d_{1x}, \ldots, d_{nx}) $ and $ D_y = (d_{1y}, \ldots, d_{ny}) $ denote two correlated attributes. For a target attribute $ x $ under attack, the correlation disruption metric $ \Delta\rho $ quantifies the impact as:  
\begin{equation}
\Delta\rho = \left| \rho_c - \rho_u \right|\label{corr_disruption_def}
\end{equation}
where $ \rho_u $ is the Pearson correlation between $ \Phi(D_x) $ and $ \Phi(D_y) $, and $ \rho_c $ is the correlation between $ \Delta(D_x) $ and $ \Phi(D_y) $.

Applying the triangle inequality and variance-covariance bounds, the maximum correlation disruption satisfies:
\vspace{-10pt}

{\small
\begin{equation}
\Delta\rho \!\leq\! {\frac{\sum_{i \in M} \left\| \left(\Delta(d_{ix}) \!-\! \Phi(d_{ix})\right)\left(\Phi(d_{iy}) \!-\! \mu_y\right) \right\|_1}{\sigma_x \sigma_y}} \!+\! {\left\|\! \frac{\sigma_x}{\tilde{\sigma}_x} \!-\! 1 \!\right\|_1} \label{corr_disruption_bound}
\end{equation}}

{where $ \mu_y = \mathbb{E}[\Phi(D_y)] $, $ \sigma_x = \text{std}(\Phi(D_x)) $, $ \tilde{\sigma}_x = \text{std}(\Delta(D_x)) $.}

By integrating the $\varepsilon$-LDP perturbation bound from (\ref{maximum_possible_deviation}), the simplified asymptotic bound is:
\begin{equation}
\Delta\rho \leq \frac{m}{n} \cdot \frac{e^\varepsilon - 1}{\sigma_x} \label{simplified_delta_rho}
\end{equation}

This reveals three critical insights for LDP poisoning impacts: (i) Tightening $\varepsilon$ (smaller $ e^\varepsilon - 1 $) linearly reduces attack effectiveness; (ii) Low-variance attributes ($\sigma_x \ll 1 $) suffer disproportionately from distortion as perturbations dominate natural variability; iii) Even small $ m/n $ ratios induce measurable correlation shifts when $ \varepsilon > 1 $.

\subsubsection{LDP Poisoning Attacks Pattern Instability \label{Consistent and Repeating Attack Pattern}}

To evade anomaly detection mechanisms, adversaries must ensure that the attack pattern remains stable and mimics the natural characteristics and inherent structure of legitimate LDP-utilized systems. Consecutive datasets $D_j^t$ and $D_j^{t-1}$ from a natural privacy-protection process typically satisfy bounded variation, a characteristic the adversary must replicate in the poisoned dataset to avoid detection:
\begin{equation}  
\left\| \Phi(D_j^{t}, \psi) - \Phi(D_j^{t-1}, \psi) \right\|_1 \leq \epsilon, \quad \forall t \label{eq:natural_stability}  
\end{equation}
with $ \epsilon> 0 $ representing maximum permissible natural drift.

To remain undetected, poisoning perturbations $ \Delta(D_j^{t}, \psi) $ must enforce stricter temporal constraints:
\begin{equation}  
\left\| \Delta(D_j^{t}, \psi) - \Delta(D_j^{t-1}, \psi) \right\|_1 \leq \epsilon_{\text{adv}} \approx \epsilon, \quad \forall t\label{eq:attack_stability}  
\end{equation}

Define the attack pattern at time instance $ t $ as:  
\begin{equation}  
P^{t} \triangleq \Delta(D_j^{t}, \psi) - \Phi(D_j^{t}, \psi) \label{eq:attack_pattern}  
\end{equation}

To achieve both stealth and impact, the adversary must simultaneously satisfy:  
\begin{align}  
\left\| P^{t} \right\|_\infty &\leq \gamma \quad \text{(Magnitude constraint)} \label{eq:attack_magnitude_constraint} \\  
\left\| P^{t} - P^{t-1} \right\|_\infty &\leq \eta \quad \text{(Variation constraint)} \label{eq:attack_variation_constraint}  
\end{align}
where $ \gamma $ limits instantaneous perturbation magnitude, and $ \eta \ll \epsilon $ enforces gradual pattern evolution.  

From \eqref{eq:natural_stability}-\eqref{eq:attack_variation_constraint}, the ratio $ \eta/\epsilon \rightarrow 0 $ becomes inevitable due to undetectability. This imposes an exponential constraint on attack adaptability:
\begin{equation}
\text{Adaptation Rate} \propto \frac{\eta}{\epsilon} \ll 1
\end{equation}

Thus, adversaries cannot rapidly adjust the attack pattern $ P^{t} $ to counter dynamic changes in the system without violating \eqref{eq:attack_variation_constraint}. Realistically, achieving perfect stability ($\eta/\epsilon\to0$) is effectively unachievable in practice. Consequently, actual poisoning vectors $P^t$ will inevitably exhibit excess temporal variation or magnitude spikes.

\section{PoisonCatcher: Identifying Data Poisoned by the LDP Poisoning Attack in IIoT}

Based on the three impacts of LDP poisoning attacks, we propose PoisonCatcher, a solution comprising three complementary time-series detectors to detect poisoned attributes (leveraging metrics like temporal similarity, attribute correlation, and pattern stability) and a latent-bias feature mining for identifying poisons.

\subsection{Temporal Similarity Detector: Detecting Anomalies from SQR Distortion}\label{Temporal Similarity Detector}

Due to process regularity, industrial process data typically exhibit inherent temporal stability. However, as demonstrated in Section \ref{Undermines the Accuracy of Statistical insights}, LDP poisoning attacks fundamentally alter this, introducing distortion in SQRs and thereby disrupting their stability. This critical contrast between the inherent stability of normal SQRs and the distorted behavior of poisoned SQRs serves as the core basis for our temporal similarity detector design. It establishes normative baselines based on the inherent temporal consistency observed in normal industrial data streams. Subsequently, it distinguishes anomalies by detecting deviations that exceed the expected variation defined by the quantifiable fault tolerance boundaries of the LDP mechanism.

\subsubsection{LDP Fault Tolerance Estimation}

The formalization of LDP fault tolerance is achieved through a dual perspective of perturbation mechanisms and statistical estimation. For continuous values, the Laplace mechanism \cite{dwork2006calibrating} is widely used for the mean estimation of LDP, while discrete values employ the generalized randomized response (GRR) \cite{christofides2003generalized} for frequency estimation. Let $\alpha$ represents maximum acceptable deviation and $\delta$ denotes confidence level, the $(\alpha, \delta)$-fault tolerance is:
\begin{equation}
\mathbb{P}\left[\left \| \mathcal{Q}(\Phi(D,\psi)) - \mathcal{Q}(D) \right \|_1 \geq \alpha \right] \leq 1-\delta
 \label{fault_tolerance}
\end{equation}

\textit{Fault Tolerance in Mean Estimation: }For attribute $j$ with continuous values employing Laplace perturbation (characterized by distribution $\mathrm{Lap}(0,|\mathcal{X}|/\varepsilon)$), we consider $n$ devices reporting values $d_{ij} \in \mathcal{X}$ with a uniform privacy budget $\varepsilon$. This analysis assumes homogeneous privacy budgets across all devices to facilitate analytical tractability, although practical implementations may require device-specific budget allocations. The resulting perturbed aggregate follows:
\begin{equation}
\mathcal{Q}(\Phi(D_j,\psi_{\mathrm{Lap}})) = \frac{1}{n}\sum_{i=1}^n \left(d_{ij} + \mathrm{Lap}\left ( \frac{2}{\varepsilon} \right ) \right)
\end{equation}

The variance of estimation error derives from Laplace mechanism properties:
\begin{equation}
\mathrm{Var}\left(\mathcal{Q}(\Phi(D_j,\psi_{\mathrm{Lap}})) - \mathcal{Q}(D_j)\right) = \frac{2|\mathcal{X}|^2}{n\varepsilon^2}
\end{equation}

Applying Chebyshev's inequality establishes probabilistic error bounds:
\begin{equation}
\mathbb{P}\left[ \left\| \mathcal{Q}(\Phi(D_j,\psi_{\mathrm{Lap}})) - \mathcal{Q}(D_j) \right\|_1 \geq \alpha_{\mathrm{Lap}} \right] \leq \frac{2|\mathcal{X}|^2}{n\alpha_{\mathrm{Lap}}^2\varepsilon^2}
\end{equation}
which illustrates how increasing the number of data or the privacy budget tightens the probabilistic guarantee, while a larger data domain range loosens it.

Solving for $\alpha_{\mathrm{Lap}}$ yields the fault tolerance boundary:
\begin{equation}\label{Lap_alpha}
\alpha_{\mathrm{Lap}} = \frac{\sqrt{2}|\mathcal{X}|}{\varepsilon\sqrt{n(1-\delta)}}
\end{equation}

\textit{Fault Tolerance in Frequency Estimation: }For the collection of discrete values, the GRR mechanism extends the randomized response (RR) to multiple categories domains \cite{christofides2003generalized}. Given a discrete value $d \in \mathcal{X}$, GRR perturbs it to $d^* \in \mathcal{X}$ with perturbation probabilities:
\begin{equation}
 \mathbb{P}\left(\Phi(d, \psi_\mathrm{GRR}) = d^*\right) = 
 \begin{cases} 
 \frac{e^\varepsilon}{e^\varepsilon + |\mathcal{X}| - 1}, & d^* = d \\
 \frac{1}{e^\varepsilon + |\mathcal{X}| - 1}, & d^* \neq d 
 \end{cases}
\end{equation}

Following the analysis framework of Kairouz et al. \cite{kairouz2016discrete}, the discrepancy between true frequencies $\mathcal{Q}(D_j)$ and GRR-perturbed estimates $\mathcal{Q}(\Phi(D_j,\psi_\mathrm{GRR}))$ in attribute $j$ satisfies:
\begin{eqnarray}
 && \left \| \mathcal{Q}(\Phi( D_j, \psi_\mathrm{(GRR)} )), \mathcal{Q}(D_j) \right \|_1 \nonumber\\
 && \approx \frac{\sum_{x=1}^{|\mathcal{X}|} \sqrt{\frac{2\left[ \left( e^{\varepsilon}-1 \right) p_{x}+1 \right] \left[ \left( e^{\varepsilon}-1 \right) \left( 1-p_{x} \right) +|\mathcal{X}|-1 \right] } {\pi n}}}{e^{\varepsilon}-1} 
\end{eqnarray}
where $p_x$ denotes the true frequency of category $x$. The estimation variance is derived as:
\begin{equation}
\mathrm{Var}\!\left(\left\| \mathcal{Q}(\Phi(D_j,\psi_\mathrm{GRR})) \!-\! \mathcal{Q}(D_j) \right\|_1\right) \!=\! \frac{4(e^\varepsilon + |\mathcal{X}| - 2)^2}{(e^\varepsilon - 1)^2\pi n}
\end{equation}

Applying Chebyshev's inequality yields the probabilistic fault tolerance bound:
\begin{equation}
    \begin{split}
     &\mathbb{P} \left[ \left \| \mathcal{Q}(\Phi( D_j, \psi_\mathrm{GRR} )), \mathcal{Q}(D_j) \right \|_1 \geq \alpha_\mathrm{GRR} \right] \\
    &\leq\frac{4\left ( e^\varepsilon +|\mathcal{X}|-2 \right )^2 }{\alpha^2_\mathrm{GRR}\pi n \left ( e^\varepsilon-1 \right )^2 }
    \end{split}
\end{equation}

Solving for $\alpha_\mathrm{GRR}$ yields the fault tolerance boundary:
\begin{equation}\label{kRR_alpha}
\alpha_\mathrm{GRR} = \frac{2(e^\varepsilon + |\mathcal{X}| - 2)}{(e^\varepsilon - 1)\sqrt{\pi n(1 - \delta)}}
\end{equation}

Equations \eqref{Lap_alpha} and \eqref{kRR_alpha} reveal three critical relationships. First, the trade-off between privacy and utility is demonstrated through $\alpha_{\mathrm{Lap}} \propto 1/\varepsilon$, establishing that increased privacy protection (lower $\varepsilon$) directly reduces the accuracy of the estimation. Second, the effect of the sample size manifested in $\alpha_{\mathrm{Lap}} \propto 1/\sqrt{n}$ confirms that the error decreases proportionally with increasing number of participants. Third, the domain dependence relationship $\alpha_{\mathrm{Lap}} \propto |\mathcal{X}|$ indicates that wider data ranges require larger tolerance thresholds to maintain accuracy.

The temporal detector monitors deviations that exceed $\alpha$ thresholds while accounting for the variation of the natural process. Sustained violations trigger poison alerts, with threshold sensitivity adjustable through parameter tuning $\delta$.

\subsubsection{Detecting Anomalies via Quantifying SQR Distortion}

The temporal similarity detector operates by establishing a dynamic baseline that integrates the inherent natural consistency of industrial data streams, as captured from historical clean datasets, with LDP fault tolerance thresholds.

The inherent natural consistency observed in industrial data streams refers to the predictable patterns, stable relationships, and statistical regularities that characterize data generated by legitimate industrial processes under normal operating conditions. These consistencies arise from the underlying physical principles that govern the industrial process, the controlled environment in which devices operate, and the designed dynamics of the components of the system. Characterizing this natural consistency is fundamental for anomaly detection, as deviations from these expected patterns serve as primary indicators of potential attacks or faults. Even in the presence of natural sensor noise or minor process fluctuations, legitimate data points adhere to these overarching consistency patterns. Modeling and understanding this inherent natural consistency using historical, unpoisoned data is therefore a crucial prerequisite for establishing a reliable baseline. 

For attribute $ j $, let $ D_j $ denote its dataset and $ \mathcal{D}_j^{\text{hist}} $ represent unpoisoned historical datasets. The similarity detection baseline is constructed as follows:
\begin{equation}
\mathcal{S}_j = \left[ \min_{D \in \mathcal{D}_j^{\text{hist}}} \mathcal{Q}(D) - \alpha, \ \max_{D \in \mathcal{D}_j^{\text{hist}}} \mathcal{Q}(D) + \alpha \right]
\end{equation}
where $ \alpha $ corresponds to the appropriate fault tolerance measure ($ \alpha_{\mathrm{Lap}} $ or $ \alpha_{\mathrm{GRR}} $) derived in \eqref{Lap_alpha} and \eqref{kRR_alpha}. The instantaneous deviation metric is computed as:
\begin{equation}
\Lambda_{\mathcal{S}_j} = \left\| \mathcal{Q}(\Delta(D_j, \psi)) - \mathcal{S}_j \right\|_1
\label{eq:temporal_similarity}
\end{equation}
\vspace{-10pt}

A value $\Lambda_{\mathcal{S}_j} > 0$ indicates a potential poisoning event and directly measures the magnitude of this deviation, with higher values corresponding to more marked anomalies.

While this temporal similarity detector effectively detects outliers, it may lack sensitivity against advanced attacks. Therefore, it should be part of a multilayered security system for thorough threat protection.

\subsection{Temporal Correlation Detector: Detecting Anomalies from Attribute Correlation Disruption}\label{Attribute Correlation Detector}

In industrial settings, the diversity and complexity of the data sources hinder the widespread poisoning, forcing adversaries to focus on vulnerable points or high-value targets \cite{mansour2020cybersecurity}. This targeted poisoning significantly disrupted previously correlated attributes, as discussed in Section \ref{Compromise the Correlation between Datasets}. Consequently, another detector is proposed for the detection of anomalies by means of an analysis of changes in correlations. This detector includes the establishment of a baseline from a clean historical clean dataset and the quantification of biases to detect subtle indications of attribute anomalies.

\subsubsection{Establishing Baselines for Mixed-Type Attribute Pairs}

The establishment of multi-attribute correlation baselines requires systematic modeling of interactions between mixed-type attributes, with particular focus on continuous and discrete values within the mean and frequency estimation. However, since LDP enforces privacy by independently perturbing each attribute (e.g.\ via per‑attribute Laplace noise or randomized response), the intrinsic correlations among original data are irreversibly masked \cite{kifer2014pufferfish,rastogi2010differentially}. Thus, we assess interattribute dependencies on the time series of SQR output of each attribute, since LDP decorrelates individual records. Tailoring the approach to the types of each attribute pair (continuous or discrete), we apply the appropriate correlation estimator, such as Pearson's correlation for continuous-continuous pairs, point-biserial correlation for continuous-discrete pairs, and sparse CCA for discrete-discrete pairs, to their SQR sequences, thereby recovering meaningful associations at the aggregate level, as detailed in Algorithm \ref{alg:correlation_analysis}.

\begin{algorithm}[htp]
\caption{Multivariate Correlation Baseline Establishment}
\label{alg:correlation_analysis}
\begin{algorithmic}[1]
\small
\Require Clean SQR dataset $\mathcal{D}$, confidence level $\delta$, bootstrap resamples $B$, window size $\ell$
\Ensure Baseline correlations $\{\hat{\rho}_{ij}\}$, confidence intervals $\{\mathcal{C}_{ij}\}$
\For{each attribute pair $(i,j) \in \mathcal{P}$}
    \If{$i$ and $j$ are continuous}
        \State Split $\mathcal{D}$ into $T-\ell+1$ windows $\{\mathcal{D}_{t:t+\ell}\}_{t=1}^{T-\ell+1}$
        \For{each window $\mathcal{D}_{t:t+\ell}$}
            \StateNoWrap {Compute Pearson correlation $\rho_{ij}^{(t:t+\ell)}$ \& sample count $n_{t:t+\ell}$}
            \State Calculate weight $w_t = \sqrt{n_{t:t+\ell}\cdot\mathrm{Var}(\rho_{ij}^{(t:t+\ell)})}$
        \EndFor
        \State $\hat{\rho}_{ij} \gets \frac{\sum_{t} w_t\rho_{ij}^{(t:t+\ell)}}{\sum_{t} w_t}$ \label{line:c2c}
    \ElsIf{$i$ is continuous, $j$ is discrete}
        \StateNoWrap {One-hot encode $j$ into $\{g_1,...,g_{j_K}\}$ with frequencies $\{n_{g_k}\}$}
        \For{each category $g_k$}
            \State Compute point-biserial correlation $\rho_{ig_k}$ 
            \State $w_k \gets \sqrt{n_{g_k}\cdot|\rho_{ig_k}|}$
        \EndFor
        \State $\hat{\rho}_{ij} \gets \frac{\sum_{k}w_k\rho_{ig_k}}{\sum_{k}w_k}$ \label{line:c2d}
    \Else \Comment{Both discrete}
        \State Solve sparse CCA:
        \State $\max\limits_{\mathbf{u}_i,\mathbf{u}_j} \mathbf{u}_i^\top\Sigma_{ij}\mathbf{u}_j - \lambda(\|\mathbf{u}_i\|_1+\|\mathbf{u}_j\|_1)$
        \State $\hat{\rho}_{ij} \gets \frac{\mathbf{u}_i^\top\Sigma_{ij}\mathbf{u}_j}{\sqrt{(\mathbf{u}_i^\top\Sigma_{ii}\mathbf{u}_i)(\mathbf{u}_j^\top\Sigma_{jj}\mathbf{u}_j)}}$ \label{line:d2d}
    \EndIf
    \State Generate $B$ bootstrap samples $\{\mathcal{D}^{(b)}\}$
    \For{each $b \in 1..B$} \Comment{Bootstrap correlations}
        \State Recompute $\rho_{ij}^{(b)}$ using same method as above
    \EndFor
    \State $\mathcal{C}_{ij} \gets \hat{\rho}_{ij} \pm Q_\delta(\{\rho_{ij}^{(b)}\}_{b=1}^B)$ \label{line:ci}
\EndFor
\State \Return $\{\hat{\rho}_{ij}, \mathcal{C}_{ij}\}_{\forall (i,j)}$
\end{algorithmic}
\end{algorithm}

\textit{Establishing Baselines for Mean Estimated Attribute Pairs: }
We apply weighted Pearson's correlation coefficients with variance stabilization to their resulting continuous-valued SQR time series. This approach compensates for temporal non-stationarity while maintaining sensitivity to persistent relationships between the mean-aggregated attributes. Given two attributes $ x $ and $ y $, the baseline correlation $ \rho_{xy} $ is computed by weighing the geometric mean of partial correlations in temporal segments:

The mean estimated SQR values of each attribute form a time series that is divided into time windows $[t : t + \ell]$ to mitigate non-stationarity. Within each time window, the Pearson correlation $\rho_{xy}^{(t:t+\ell)}$ is computed; its variance $\mathrm{Var}(\rho_{xy}^{(t:t+\ell)})$ and sample count $n_{t:t+\ell}$ define a weight:
\begin{equation}
    w_{t:t+\ell} = \sqrt{\,n_{t:t+\ell}\,\mathrm{Var}\bigl(\rho_{xy}^{(t:t+\ell)}\bigr)}
\end{equation}

The baseline correlation then is the weighted average:
\begin{equation}
    \rho_{xy}=\frac{\sum_{t=1}^{T-\ell+1} w_{t:t+\ell}\,\rho_{xy}^{(t:t+\ell)}}
{\sum_{t=1}^{T-\ell+1} w_{t:t+\ell}}\label{c2c}
\end{equation}
which balances temporal drift against the need to remain sensitive to enduring linear relationships.

\textit{Establishing Baselines for Mixed-Estimation Attribute Pairs: }For pairs comprising a scalar mean SQR and a vector frequency SQR, their correlation is modeled using hierarchical regression composite correlation. To quantify the correlation between the attribute $x$ and $y$, let $j_K$ denote the number of categories of $y$, $\rho_{xK}$ the point‑biserial correlation between $x$ and the $K$th frequency category of $y$, and $n_{y_K}$ the frequency of the corresponding category. We assign each category a weight: 
\begin{equation}
w_K = \sqrt{\,n_{y_K}\,\bigl|\rho_{xK}\bigr|\,}
\end{equation}
and define the baseline correlation as:
\begin{equation}
  \rho_{xy} = \frac{\sum_{K=1}^{y_K} w_K\,\rho_{xK}} {\sum_{K=1}^{y_K} w_K}\label{c2d}
\end{equation}

By boosting well-populated, strongly correlated categories, this approach improves estimator robustness and accuracy under LDP noise and sparse distributions.

\textit{Establishing Baselines for Frequency Estimated Attribute Pairs}: The frequency-based SQR distributions from pairs are analyzed using $\ell_1$ regularized canonical correlation analysis (CCA) to isolate their most significant covariation patterns. This sparse CCA framework not only highlights the significant correlation patterns, but also suppresses the noise and redundant features, improving the interpretability and robustness of the model in high-dimensional sparse environments.

Let $\Sigma_{xx}$, $\Sigma_{yy}$ be the covariance matrices within the attribute, and $\Sigma_{xy}$ the cross-covariance. The optimal sparse projections $\mathbf{u}_x^*\in\mathbb{R}^{x_K}$, $\mathbf{u}_y^*\in\mathbb{R}^{y_K}$ solve:
\begin{equation}
(\mathbf{u}_x^*,\mathbf{u}_y^*)=\arg\max_{\mathbf{u}_x,\mathbf{u}_y}\,
\mathbf{u}_x^\top\Sigma_{xy}\mathbf{u}_y - \lambda\bigl(\|\mathbf{u}_x\|_1 + \|\mathbf{u}_y\|_1\bigr)
\end{equation}
where $ \lambda $ controls sparsity. The resulting correlation is:
\begin{equation}
\rho_{xy}=\frac{\mathbf{u}_x^{*\top}\Sigma_{xy}\mathbf{u}_y^*}
{\sqrt{\bigl(\mathbf{u}_x^{*\top}\Sigma_{xx}\mathbf{u}_x^*\bigr)\,\bigl(\mathbf{u}_y^{*\top}\Sigma_{yy}\mathbf{u}_y^*\bigr)}}\label{d2d}
\end{equation}

The confidence intervals of attribute pair $(x,y)$ are constructed through nonparametric residual bootstrapping:
\begin{equation}
\mathcal{C}_{xy} = \hat{\rho}_{xy} \pm Q_\delta\left( \{\rho_{xy}^{(b)}\}_{b=1}^{B} \right)\label{confidence intervals of attribute pair}
\end{equation}
where $\hat{\rho}_{xy}$ denotes the baseline correlation coefficient, estimated from historical SQR time‑series under normal operating conditions; $Q_\delta$ denotes the $\delta$ empirical quantile of bootstrapped correlations, and $B$ represents the number of bootstrap resamples. This dynamic formulation accounts for operational regime transitions and measurement uncertainty.

\subsubsection{Detecting Anomalies via Quantifying Attribute Correlation Disruption}

To detect faint signals of attribute anomalies, we introduce a pairwise to attribute bias quantification framework. For each attribute $j$, we compute the cumulative deviation:
\begin{equation}  
\Delta\rho_j = \sum_{x \in \mathcal{P}_j} |\rho_{xj}^{\text{obs}} - \hat{\rho}_{xj}|  
\end{equation}
where $\rho_{xj}^{\text{obs}}$ is the observed correlation between attributes $x$ and $j$; $\hat{\rho}_{xj}$ is the expected correlation between attributes $x$ and $j$, derived from the baseline \eqref{c2c}\eqref{c2d}\eqref{d2d}; $\mathcal{P}_j \subset \mathcal{P}$ is the subset of pairs involving attribute $j$. The sum $\Delta\rho_j$ thus quantifies the cumulative deviation of observed relationships involving attribute $j$ from their expected values.

The boundary violation metric $\Lambda_{\mathcal{C}_j}$ then quantifies the minimal distance to historical baselines:
\begin{equation}  
\Lambda_{\mathcal{C}_j} = \min\left( |\Delta\rho_j - \mathcal{C}_{j}^L|, |\Delta\rho_j - \mathcal{C}_{j}^U| \right)  \label{boundary violation metric}
\end{equation}
with $\mathcal{C}_{j}^{L/U}$ being the confidence bounds from \eqref{confidence intervals of attribute pair} for attribute $j$. The anomalous attribute is identified as:
\begin{equation}  
\hat{j} = \mathop{\arg\max}\limits_{j \in \mathcal{A}} \; \Lambda_{\mathcal{C}_j}  
\end{equation}
where $\mathcal{A}$ denotes the monitorable attribute set. This equation selects the attribute with the maximum baseline divergence, prioritizing the critical infrastructure components that show the strongest attack signatures.

This multifaceted approach offers a robust defense against abrupt correlation shifts. However, sophisticated adversaries may attempt gradual correlation deviation, necessitating integration with other temporal detectors for comprehensive protection.

\subsection{Time-Series Stability Detector: Detecting Poisoned Attributes from Attack Pattern Variation}

As established in Section \ref{Consistent and Repeating Attack Pattern}, poisoning streams hardly exhibit inherent temporal instability. To exploit this, we design a time-series stability detector that ingests the quantified temporal biases produced by the similarity and correlation detector modules (\ref{Temporal Similarity Detector} and \ref{Attribute Correlation Detector}). Within each observation window, stability metrics are computed, and any attribute whose bias exceeds its empirical threshold in these metrics is flagged as poisoning to distinguish sustained attack patterns from transient system or noise-induced variations.

Let $[t,t+\ell]$ denote a discrete time‐series observation window.  For each attribute $j$, define the similarity‐bias and correlation‐bias sequences as:
\begin{equation}
\{\Lambda_{x_j}^\tau\}_{\tau = t}^{t+\ell} = \left[ \Lambda_{x_j}^t, \Lambda_{x_j}^{t+1}, \ldots, \Lambda_{x_j}^{t+\ell} \right], x \in \{\mathcal{S},\mathcal{C}\}
\end{equation}
where $\Lambda_{x_j}^\tau$ is computed respectively from \eqref{eq:temporal_similarity} ($x=\mathcal S$) and from \eqref{boundary violation metric} ($x=\mathcal C$).

The stability of temporal sequences is quantified through three complementary metrics. Statistical dispersion is evaluated via the sequence variance $\text{Var}(\{\Lambda_{x_j}^\tau\}_{\tau = t}^{t+\ell})$, which captures deviations from the temporal mean. 

Dynamic range analysis employs the maximum fluctuation amplitude: 
\begin{equation}
R(\{\Lambda_{x_j}^\tau\}_{\tau = t}^{t+\ell}) = \max_{\tau,\tau' \in [t,t+\ell]} \left| \Lambda_{x_j}^\tau - \Lambda_{x_j}^{\tau'} \right|
\end{equation}
quantifying extreme variations within the observation window.

Temporal dependency is defined by the first-order autocorrelation coefficient:
\begin{equation}
\rho^{\text{auto}}(\{\Lambda_{x_j}^\tau\}_{\tau = t}^{t+\ell}) = \frac{\sum_{\tau=t}^{t+\ell-1} (\Lambda_{x_j}^\tau - \mu_{x_j})(\Lambda_{x_j}^{\tau+1} - \mu_{x_j})}{\sum_{\tau=t}^{t+\ell} (\Lambda_{x_j}^\tau - \mu_{x_j})^2}
\end{equation}
where $\mu_{x_j}$ denotes the temporal mean of $\{\Lambda_{x_j}^\tau\}_{\tau = t}^{t+\ell}$. 

These metrics collectively evaluate volatility, extreme behavior, and the temporal persistence of bias patterns, thereby defining a stability measure space:
{\small
\begin{equation}
\mathcal{M}_x = \left\{ \text{Var}(\{\Lambda_{x_j}^\tau\}_{\tau = t}^{t+\ell}), R(\{\Lambda_{x_j}^\tau\}_{\tau = t}^{t+\ell}),  \left|\rho^{\text{auto}}(\{\Lambda_{x_j}^\tau\}_{\tau = t}^{t+\ell})\right| \right\}
\end{equation}}

Let $\mathcal{M}_x[i]$ denote the $i$-th metric in $\mathcal{M}_x$. A robust hypothesis testing framework demands simultaneous threshold breaches in both similarity and correlation dimensions:
\begin{equation}
\begin{cases} 
\mathcal{H}_1: \exists i \in \{1,2,3\}, \, \mathcal{M}_{\mathcal{S}}[i] > \theta_i \land \mathcal{M}_{\mathcal{C}}[i] > \theta_i \\ 
\mathcal{H}_0: \text{Otherwise} 
\end{cases}
\end{equation}
where thresholds $\{\theta_i\}_{i=1}^3$ are derived empirically through systematic analysis of historical benign data.

The integration of temporal similarity, attribute correlation, and pattern stability analyzes enables the flagging of attributes only when these three criteria are violated, thereby ensuring high sensitivity to poisoning while minimizing false positives.

\subsection{Latent-Bias Features Miner: Identifying Poison from Subtle Poisoning Signatures}

LDP adds noise to individual data, weakening the ability of traditional anomaly detection methods to identify poisons (see Section \ref{Poisoned Data Points are Stealthy}). To address this limitation, we propose a latent-bias features miner, which consists of an enhanced feature engineering (FE-enhanced) approach to obtain indicative features of spatio-temporal biases, and a machine learning model to identify poisoned data.

\subsubsection{FE-enhanced Approach}

Random sampling, as a fundamental statistical tool, is well-suited to investigate data distributions and detect rare patterns within a dataset. In our approach, repeated random sampling and subsequent aggregation are applied to the time-series attribute dataset of detected poisoned to systematically amplify the subtle statistical biases induced by data poisoning, which are often not readily apparent at a single time instance. This enhances the detection of latent bias features, providing a strong basis for identifying poison.

LDP poisoning attacks typically introduce systematic bias in the statistical distributions of the targeted attributes (see Section \ref{Undermines the Accuracy of Statistical insights}). Despite LDP obscuring individual biases, aggregate statistical bias still serves as a detectable signal. The FE-enhanced approach focuses on identifying and leveraging the statistical indicators that are sensitive to such poisoning. Candidate features are selected for their theoretical sensitivity to attack-induced changes in statistical properties, like shifts in central tendencies (mean, median), dispersion (variance), and distributional shape (e.g., KL divergence).

A systematic sampling and aggregation procedure is used to extract discriminative features that capture the dynamic evolution of bias over time. This extraction procedure is necessary because subtle spatio-temporal poisoning bias patterns, which are often not readily apparent in raw data representations, require dedicated processing to be effectively captured and highlighted for subsequent detection.

At each time instance $t$ and for each device $i$, we perform $s$ independent random sampling operations, computing the bias feature for each sub-sample. An aggregation function (e.g. the arithmetic mean) is then applied to fuse the $s$ observations into a scalar value $F_i^t$, representing the aggregated statistical state of the device $i$ at the time instance $t$. Collecting the values $F_i^t$ for all devices $n$ at the time instance $t$ forms a state vector of the system of $n$ dimensions $\mathbf{F}^t = [F_1^t, F_2^t, \ldots, F_n^t]$. Finally, spanning a temporal window of length $\ell$, we sequence the $n$-dimensional state vectors $\mathbf{F}^{t}$ as columns to construct the $n \times \ell$ feature matrix $\mathbf{F}^{t:t+\ell}$. This matrix encapsulates the spatio-temporal evolution of the bias features and serves as input for the subsequent poison identification model.

\subsubsection{ML-Based Poison Identification}

To achieve an accurate identification of the poisoning status of data sources, we use a supervised classification framework that ingests the bias feature matrix $\mathbf{F}^{t:t+\ell}$. This FE-enhanced approach can integrate various classification algorithms, including logistic regression, random forest classifier, support vector machines, etc., customized to specific application scenarios and data characteristics.

Taking the random forest (RF) classifier as an illustrative example, the spatio-temporal bias feature matrix $\mathbf{F}^{t:t+\ell} \in \mathbb{R}^{n \times \ell}$, where each row $\mathbf{f}_i = \mathbf{F}\left[i, :\right] \in \mathbb{R}^\ell$ represents the bias features of the device $i$ across the temporal window, serves as input to the model. The RF classifier processes each vector $\mathbf{f}_i$ independently, applying a learned function $f_{\text{RF}}$ to output a binary poisoning label $\hat{y}_i \in \{0, 1\}$ for device $i$, defined as:
\begin{equation}
    \hat{y}_i = f_{\text{RF}}(\mathbf{f}_i)
\end{equation}
where, $\hat{y}_i=1$ indicates that device $i$ is identified as poisoned and $\hat{y}_i=0$ as unpoisoned. 

Suitable classification algorithms for this task must handle the high-dimensional $\mathbb{R}^\ell$ feature space and exhibit robustness to noise, particularly against stochastic perturbations introduced by LDP. The RF classifier, as an illustrative example, exhibits significant promise due to their ensemble learning and randomization strategies that provide such capabilities. During training, a supervised classifier learns the mapping function from $\mathbb{R}^\ell$ to $\{0, 1\}$. This enables effective discrimination between poisoned and unpoisoned LDP-protected data, providing reliable automated identification. Our approach is flexible and supports the integration of any supervised classification algorithm that meets these requirements.

\section{Experimental Analysis}

PoisonCatcher is implemented in Python 3.9. All experiments are carried out on a Windows 11 machine with a 13th Gen Intel® Core™ i7-13700 2.10 GHz and 16 GB of main memory. To promote the replicability of the research, the complete code implementation has been publicly released in the github repository \cite{li2024poisoncatcher}.

\subsection{Experimental Setup}

\textit{Dataset}: PoisonCatcher was evaluated on the World Weather Repository \cite{global_weather_repository_kaggle}.  From the original data, we first removed any samples exhibiting non-uniform spatial coverage or irregular temporal sampling.  The remaining numerical attributes were then rescaled by Min-Max normalization to the interval $[-1,1]$, ensuring consistent feature magnitudes.  Next, we discarded variables unsuitable for mean or frequency-based analysis, specifically those with excessive dispersion or pronounced skewness.  The final benchmark set comprises 56 countries observed over 288 consecutive days, each described by 10 standardized attributes.

\textit{Performance Metrics}: The F2 score was adopted as the primary quantitative metric to evaluate PoisonCatcher performance in both detecting poisoned attributes and identifying poisoned data with varying intensities of poisoning. This decision is rooted in the critical asymmetric costs inherent in IIoT poisoning detection, where the severe impact of false negatives mandates a significantly higher emphasis on recall. By weighting recall twice as heavily as precision , the F2 score mathematically captures this priority, providing a singular and relevant measure.

\textit{Parameter Settings}: Unless stated otherwise, both the GRR and Laplace mechanisms use a privacy budget of 1 and a confidence level of 95\%. These values are based on standard LDP principles, where a balance is struck between privacy and utility, and the confidence level of 95\% provides statistical reliability without imposing excessive computational demands. The three attack modes are carefully designed to balance stealth and disruptiveness, ensuring that the intended attack effects are achieved while remaining difficult to detect. Specifically, in DIPA, the data deviation is set to achieve the maximum allowable deviation defined by \eqref{stealth_1}; DRPA applies random privacy budgets within the range specified by \eqref{stealth_2}; and ROPA operates within the random offset values constrained by \eqref{stealth_3}.

\subsection{Experimental Results and Analysis}

\subsubsection{Efficacy of PoisonCatcher's Triple-Detector for Poisoned Attribute Detection}

To evaluate PoisonCatcher's effectiveness in detecting poisoned attributes, this experiment assesses Triple-Detector sensitivity and robustness by observing F2 score variations across attack ratios from 0\% to 50\%.

\begin{figure*}[htbp]
\centering
\includegraphics[width=\linewidth]{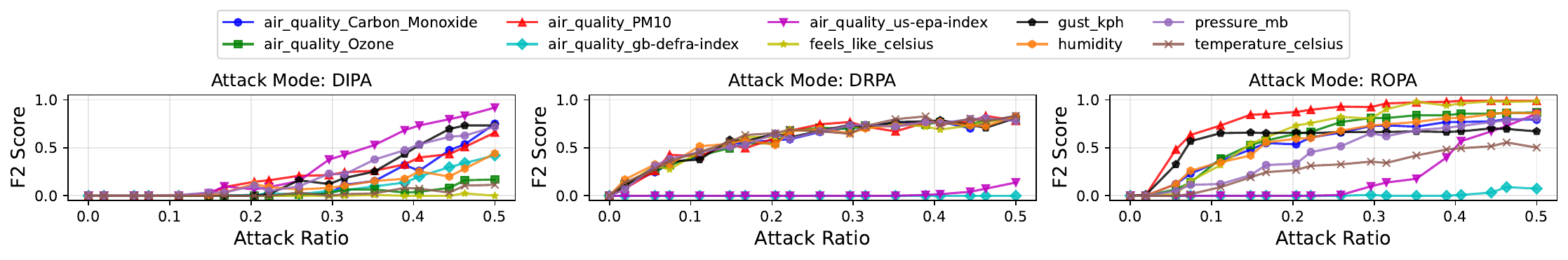}
\caption{Efficacy of Temporal Similarity Detector in Detecting Poisoned Attributes}
\label{Experiment_1}
\end{figure*}

\textit{Exp 1. Efficacy Assessment of the Temporal Similarity Detector: } This experiment evaluates the detector's ability to detect poisoned attributes, with results shown in Fig. \ref{Experiment_1}.
 
\textit{The Detector for the DIPA Experiment} showed a significant detection lag. Specifically, the F2 score remains consistently below 0.25 at attack ratio below 25\%, but begins to increase gradually as the ratio increases beyond 25\%. This lag effect is likely due to the limited variability in the sensor sensing range, which makes it statistically challenging to differentiate poisoned samples from normal ones.

\textit{The Detector for the DRPA Result} demonstrated that the F2 score for continuous-value attributes increased steadily with the attack ratio. Conversely, the detector for discrete-value attributes struggled to detect poisoning. This phenomenon arises from the inherent characteristics of different attribute distributions and their varying sensitivities to targeted bias. Attributes with narrow ranges or high volatility tend to reveal privacy budget manipulations more easily, particularly those with continuous-values normalized within the range [-1,1].

\textit{The Detector for the ROPA Experiment Result} showed that the F2 score for all attributes increased with the attack ratio, although the detection efficacy was notably insufficient at low attack ratios. This results from ROPA's exploitation of LDP's tolerance to noise, thereby masking early signs of poisoning.

The variation in F2 score curves between the attributes in the three attack experiments is due to the unique distribution of each attribute. When an attribute distribution exhibits more unimodal, concentrated, and symmetric characteristics, the statistical estimate of mean and frequency will more accurately reflect the central tendency, and the statistical deviations remain within the threshold even after the introduction of LDP noise, making anomalous patterns easier to detect.

In summary, the Temporal Similarity Detector performs inadequately at lower attack ratios, specially when detecting poisoned attributes in DIPA, highlighting the need for more detectors to improve detection accuracy.

\begin{figure*}[htbp!]
\centering
\includegraphics[width=\linewidth]{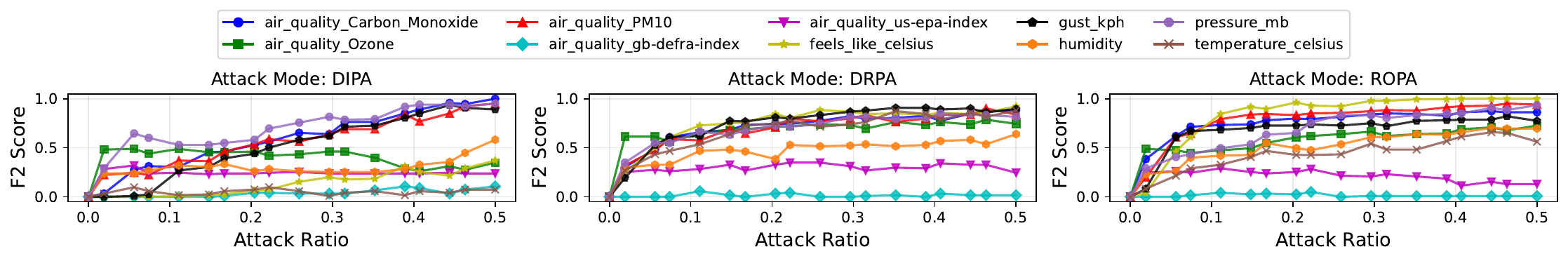}
\caption{Efficacy of Attribute Correlation Detector in Detecting Poisoned Attributes}
\label{Experiment_2}
\end{figure*}

\textit{Exp 2. Efficacy Assessment of the Attribute Correlation Detector:} The efficacy of the detector in detecting poisoned attributes is evaluated in this experiment, and the corresponding results are depicted in Fig. \ref{Experiment_2}.

In the detection of DIPA, the attribute correlation detector outperforms the temporal similarity detector. However, for DRPA and ROPA, both detectors exhibit comparable performance, with the attribute correlation detector showing a slight advantage at lower attack ratios. Notably, the attribute correlation detector exhibits complete detection failure when applied to discrete-valued attributes, demonstrating its fundamental incompatibility with non-continuous data structures. This result attributed to the limited range of discrete values, minor variations in frequency distributions may not be sufficiently captured by the attribute correlation threshold, rendering the poison introduced by the attack undetectable.

Individually, the Attribute Correlation Detector demonstrates inadequate performance. Furthermore, its combination with the Temporal Similarity Detector offers only a limited improvement in detection accuracy. Thus, enhancing the detection accuracy of poisoning characteristics necessitates the deployment of additional detectors.

\begin{figure*}[htbp!]
\centering
\includegraphics[width=\linewidth]{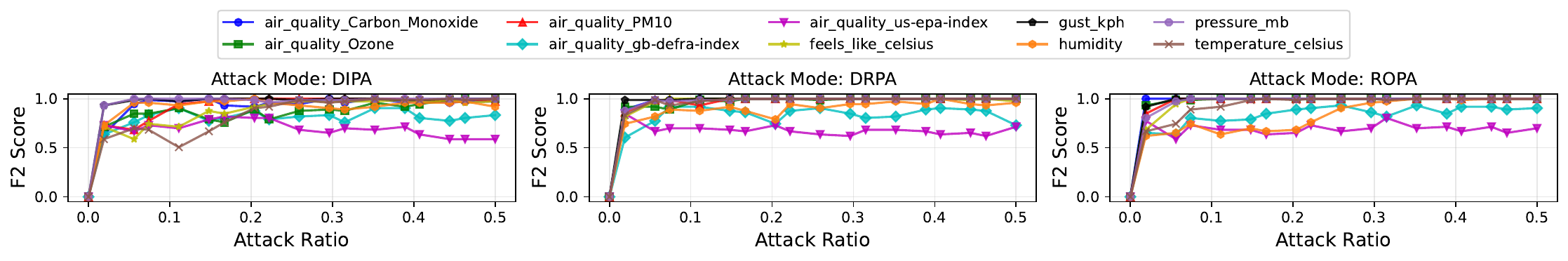}
\caption{Efficacy of Stability Tracking Detector in Detecting Poisoned Attributes}
\label{Experiment_3}
\end{figure*}

\textit{Exp 3. Efficacy Assessment of the Stability Tracking Detector:} The detector evaluates the temporal stability of bias from the outputs of the two preceding detectors, by analyzing variance, maximum fluctuation amplitude and first-order autocorrelation coefficients, as illustrated in Fig. \ref{Experiment_3}.

The experimental results against DIPA demonstrated that the detector successfully identified poisoned attributes at the attack ratio 3\%. Five attributes consistently exhibited high F2 scores (95.3\%), while another five displayed F2 scores that fluctuated around 76.2\%. For DRPA, the F2 scores for continuous value attributes reached approximately 91.6\% in the attack ratio of 8\%, increasing to approximately 98.6\% in the attack ratio of 13\%. In addition, the discrete value attributes maintained an average F2 score of approximately 73.9\%. Against ROPA, the F2 scores for continuous value attributes exceeded 98.1\% in an attack ratio 13\%, while discrete attributes maintained an average F2 score of approximately 73\%.

In summary, although the first two detectors exhibit limited effectiveness, the time-series stability detector significantly improves detection accuracy. Detecting DRPA and ROPA, the attack ratio of 5\% yields an F2 score exceeding 98\% for continuous-value attributes and approximately 73.5\% for discrete-value attributes. Detecting DIPA, despite some fluctuations, the average F2 score for continuous and discrete value attributes is around 75\%. These results demonstrate that the three detectors, when used collectively (despite individual limitations), are highly effective in detecting poisoned attributes.

\subsubsection{Efficacy of PoisonCatcher for Attack Ratio Estimation}

Current research on LDP poisoning attacks focuses mainly on defense mechanisms, with little focus on identifying specific poisoned data. To date, only DETECT \cite{272214} and LDPGuard \cite{10415225} have been proposed to estimate the poisoning ratio in the datasets. In contrast, PoisonCatcher is the first solution that identifies poisoned data. By precisely locating poisoned samples instead of simply estimating a global poisoning ratio, PoisonCatcher enables more fine-grained detection.

To demonstrate the effectiveness and advancement of PoisonCatcher, we designed a multi-dimensional experimental setup: (i) using PoisonCatcher’s identification accuracy as an indirect estimate of the poisoning ratio; (ii) comparing the direct poisoning ratio estimates provided by DETECT \cite{272214} and LDPGuard \cite{10415225}; (iii) construct a theoretical poisoning intensity estimate based on \eqref{eq:accuracy_degradation} and \eqref{simplified_delta_rho}; (iv) incorporate an expert evaluation group comprising eight information security experts from authoritative institutions (including three AI-based evaluation models) to perform poisoning ratio estimation; and (v) conduct ablation studies to compare the effectiveness of a standard RF classifier with that of a model that combines FE-enhanced and RF classifier to estimate poisoning ratio. The experimental results are shown in Table \ref{tab:attack_success}.

\begin{table*}[ht]
\centering
\caption{Comparison of Attack Ratio Estimates}
\label{tab:attack_success}
\renewcommand{\arraystretch}{0.9}
\begin{tabular}{@{}ccccccc@{}cc}
\toprule
\textbf{Protocols}& \textbf{Attack Mode}&   \makecell{\textbf{True Attack} \\ \textbf{Ratio}} & \makecell{\textbf{DETECT \cite{272214}}\\ \textbf{Estimate}} & \makecell{\textbf{Expert} \\ \textbf{Estimate}} &\makecell{\textbf{LDPGuard} \cite{10415225} \\ \textbf{Estimate}} & \makecell{\textbf{ Theoretical  } \\ \textbf{ Estimation  }} &\makecell{\textbf{Random Forest} \\ \textbf{Identification}}& \makecell{ \textbf{PoisonCatcher}\\ \textbf{Identification}}\\
\midrule
\multirow{3}{*}{Laplace} & DIPA &   5\%
&— &3.46\%&—& 1.15\%   &4.85\%&4.98\%\\
 & DRPA &   5\%
&— &8.66\%&—& 20.78\%  &4.88\%&5\%\\
 & ROPA &   5\%
&— &9.35\%&—& 11.65\%  &4.86\%&5\%\\
\midrule
\multirow{3}{*}{KRR} & DIPA &   5\%
&— &5.68\%&59.30\% & 3.46\%   &4.83\%&4.97\%\\
 & DRPA &   5\%
&— &6.72\%&—& 1.2\%  &4.88\%&4.99\%\\
 & ROPA &   5\%&— &8.62\%&38.20\% & 1.03\%  &4.87\%&5\%\\
\bottomrule
\end{tabular}
\end{table*}

The experimental results indicate that, although PoisonCatcher's attack ratio estimates are not perfectly precise, they more closely approximate the true values than those of alternative methods, resulting in smaller errors. Furthermore, the DETECT method is completely ineffective in this scenario because it identifies potential fake users solely based on their reported values. These findings suggest that PoisonCatcher exhibits notable effectiveness and advances in estimating poisoning attack ratios.

\subsubsection{Efficacy of PoisonCatcher for Identification Poisoned Data}

{To evaluate the efficacy of PoisonCatcher in identifying poisoned data, we compared the performance of two models: a baseline RF classifier and an RF classifier incorporating our proposed FE-enhanced approach. Performance was assessed using the F2 score, measured in poisoning ratios ranging from 0\% to 50\%.}

\begin{figure*}[htbp!]
\centering
\includegraphics[width=\linewidth]{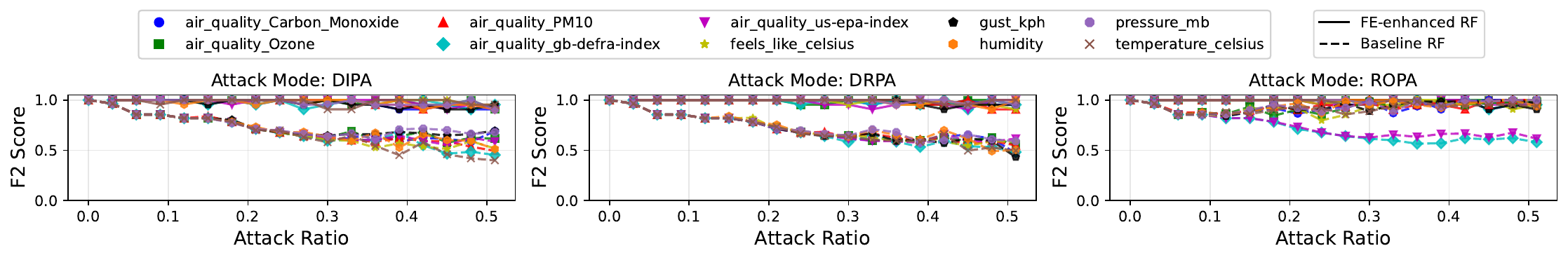}
\caption{Efficacy of Attack Ratio on Per-Attribute Poisoned Data Identification Performance}
\label{Experiment_5}
\end{figure*}

The results in \ref{Experiment_5} show the effectiveness of RF with the FE-enhanced approach in improving the performance of poisoning identification. Compared to RF only, which exhibits a significant decline in the F2 score of up to 43.2\% with increasing attack ratios, RF with FE-enhanced approach consistently maintains a stable identification performance (F2 score $\geq$ 90.7\%) across attack ratios ranging from 0\% to 50\% and three attack modes. Furthermore, evaluations in nine bias features (include mean, median, variance, MAE, KL, SQR bias, hypothesis testing bias in stratified or unstratified sampling, individual variance) indicate that each feature achieves a consistently high identification accuracy (F2 score $\geq$ 90.2\%), further confirming the robustness of the RF with FE-enhanced approach. Overall, these findings underscore the critical role of FE-enhanced approach in achieving robust and stable poisoning identification across varying attack conditions. Due to similar results, no further figures will be provided here.

\subsubsection{Robustness of PoisonCatcher for Identification Poisoned Data}

Beyond evaluating overall identification efficacy, we further assessed the robustness of our proposed latent-bias feature miner by investigating the performance stability of both the baseline RF models and FE-enhanced RF models under varying LDP privacy budgets and time instance lengths, specifically at a 5\% attack ratio.

\begin{figure*}[htbp!]
\centering
\includegraphics[width=\linewidth]{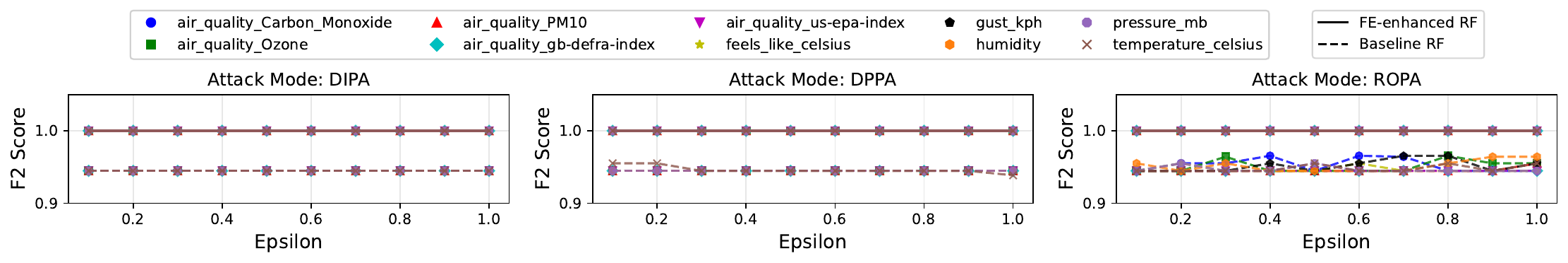}
\caption{Robustness of Privacy Budget on Per-Attribute Poisoned Data Identification Performance}
\label{Experiment_6_1}
\end{figure*}

\begin{figure*}[htbp!]
\centering
\includegraphics[width=\linewidth]{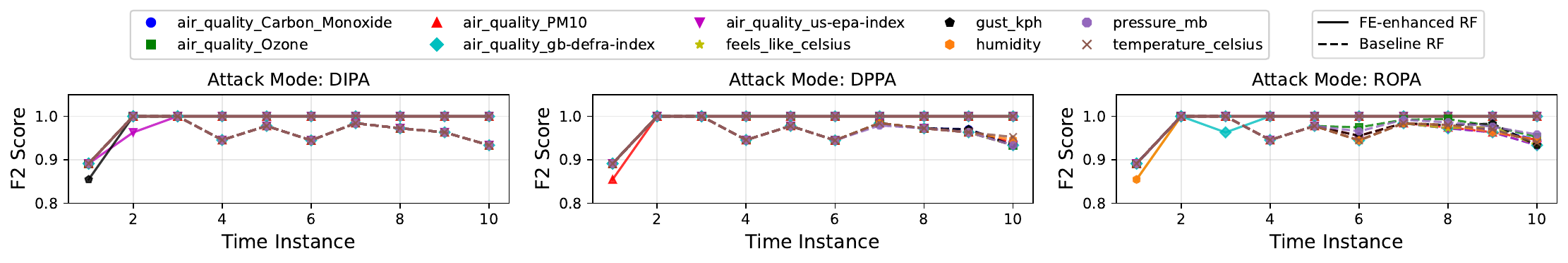}
\caption{Robustness of Length of Time Window on Per-Attribute Poisoned Data Identification Performance}
\label{Experiment_6_2}
\end{figure*}

As depicted in Fig. \ref{Experiment_6_1} and Fig. \ref{Experiment_6_2}, the FE-enhanced with RF consistently achieves near-optimal (F2 score close to 0.993) and stable performance across all attributes, privacy budgets, and attack modes, and rapidly converges by time window greater than 2. In contrast, baseline RF exhibits lower overall efficacy, significant temporal instability, and high performance variability that is strongly dependent on the specific data attribute, particularly in the ROPA attack scenario.

In summary, the FE-enhanced approach not only ensures robust, high-fidelity identification but also delivers rapid convergence, enabling reliable real-time identification in LDP-utilized IIoT environments.

\section{Conclusions}

{This work reveals a critical vulnerability in LDP-utilized IIoT systems, where adversaries can exploit the inherent indistinguishability of LDP and the spatio-temporal complexity of IIoT to execute stealthy LDP poisoning attacks, thereby compromising data utility. Through the analysis of various attack scenarios, their impacts and charicteristics, we developed PoisonCatcher, a four-stage solution that consistently achieves an F2 score of at least 90.7\% in identifying poisoned data across all attack intensities and distinct attack modes on real-world datasets.}

{Interesting future work includes generalizing PoisonCatcher in two complementary directions. First, we will extend our evaluation to diverse, publicly available IIoT datasets to further validate the real-world applicability and generalizability of the solution. Second, we will adapt and deploy PoisonCatcher in other privacy-preserving contexts, beyond the LDP setting, to defend against poisoning attacks in a wider range of protected data‐sharing scenarios. Together, these efforts will not only reinforce the practical utility of PoisonCatcher but also broaden its applicability across evolving IIoT and privacy-sensitive environments.}


\section*{Acknowledgement}
The authors thank anonymous reviewers for their constructive suggestions, which significantly improved the quality of this manuscript. We also thank Qi Shi for their invaluable assistance in acquiring the real-world dataset used in the experiments. 



%
\renewcommand*{\bibfont}{\footnotesize} 
 \printbibliography{}

\end{document}